# Magnetic phase diagram of the triangular antiferromagnetic Cs$_2$CuCl$_{4-x}$Br$_x$ mixed system

*Natalija van Well,* *Oksana Zaharko, Bernard Delley, Markos Skoulatos, Robert Georgii, Sander van Smaalen, Christian Rüegg*

Dr. N. van Well
Laboratory for Neutron Scattering and Imaging, Paul Scherrer Institute, CH-5232 Villigen, Switzerland
Laboratory of Crystallography, University of Bayreuth, 95447 Bayreuth, Germany
E-mail: Natalija.van-well@uni-bayreuth.de
Dr. O. Zaharko
Laboratory for Neutron Scattering and Imaging, Paul Scherrer Institute, CH-5232 Villigen, Switzerland
Dr. B. Delley
Condensed Matter Theory Group, Paul Scherrer Institute, CH-5232 Villigen, Switzerland
Dr. M. Skoulatos
Heinz Maier-Leibnitz Zentrum (MLZ), Technische Universität München, Lichtenbergstr. 1, 85747 Garching, Germany
Physik Department E21, Technische Universität München, James-Franck-Str., 85747 Garching, Germany
Dr. R. Georgii
Heinz Maier-Leibnitz Zentrum (MLZ), Technische Universität München, Lichtenbergstr. 1, 85747 Garching, Germany
Physik Department E21, Technische Universität München, James-Franck-Str., 85747 Garching, Germany
Prof. Dr. S. van Smaalen
Laboratory of Crystallography, University of Bayreuth, 95447 Bayreuth, Germany
Prof. Dr. Ch. Rüegg
Laboratory for Neutron Scattering and Imaging, Paul Scherrer Institute, CH-5232 Villigen, Switzerland
Department for Quantum Matter Physics, University of Geneva, CH-1211 Geneva, Switzerland

The novel magnetic phase diagram of the Cs$_2$CuCl$_{4-x}$Br$_x$ mixed system is established by means of single crystal neutron diffraction in the lowest temperature region and zero magnetic field. Two long-range ordered magnetic phases exist in this mixed system depending on the Cl/Br concentration. In the rich Cl concentration range, the ordered magnetic state occurs below the ordering temperature $T_N = 0.51(1)$K for $Cs_2CuCl_3Br_1$ and at $Cs_2CuCl_{2.6}Br_{1.4}$ below $T_N = 0.24(2)$K. Magnetic order with a temperature-independent position $(0, 0.573(1), 0)$ below the ordering temperature $T_N = 0.63(1)$K appears in the rich Br concentration for $Cs_2CuCl_{0.6}Br_{3.4}$. Between the rich Cl and rich Br concentration ranges (two magnetic phases), there is a range of x without magnetic order down to 50mK. A suggestion about the magnetic exchange paths in the bc-layer for different regimes is presented, which can be controlled depending on the preferred Br-occupation in the [CuX$_4$] tetrahedra. The density functional theory (DFT) calculations of the exchange coupling constants $J$, $J'$ for some ordered compositions of the mixed system $Cs_2CuCl_{4-x}Br_x$ indicate that these are not frustrated.

## 1. Introduction

Quantum antiferromagnets in low dimensions have been a central concern in the last decades.[1-3] $Cs_2CuCl_4$ and $Cs_2CuBr_4$ are quasi-two dimensional (2D) frustrated quantum antiferromagnets.[4,5] They are model anisotropic triangular lattice materials, where [Cu$^{2+}$] ions with $S = ½$ form chains with magnetic exchange coupling $J$, which are zig-zag coupled with $J'$ to form a frustrated 2D lattice on the bc-plane.[4-6] Neighbouring layers are coupled by a small exchange $J''$ in a-direction (**Figure 1**).

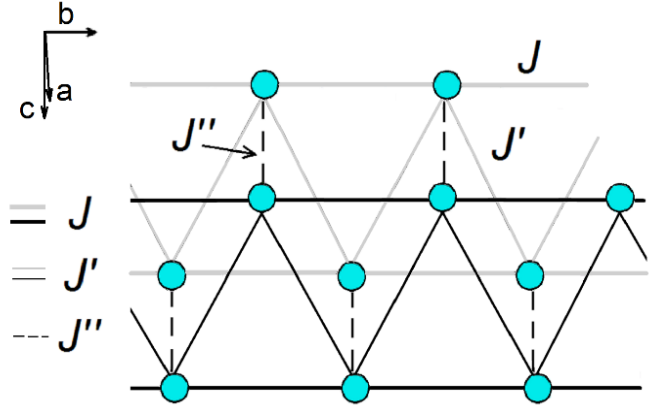

**Figure 1.** Schematic magnetic exchange interactions between [Cu$^{2+}$] ions within the bc-plane.

For $Cs_2CuCl_4$ and $Cs_2CuBr_4$, $J$ and $J'$ are both antiferromagnetic. Dominant exchange interactions are along the chains (along the b-axis), with $J'/J=0.34$ for $Cs_2CuCl_4$ and $J'/J=0.74$ for $Cs_2CuBr_4$.[5,7] For $Cs_2CuCl_4$ an antiferromagnetic (AFM) order exists below $T_N = 0.62$ K.[6] A quantum phase transition occurs at a magnetic field along the a-direction of $H_C = 8.44$ T, where in higher fields the order is ferromagnetic (FM).[8] High-resolution time-of-flight neutron spectroscopy revealed magnetic excitations interpreted as 2D spin liquid phase above the magnetic ordering transition.[8] The dynamical correlations are dominated by highly dispersive scattering continua, characteristic for the fractionalization of spin waves into pairs of deconfined $S = ½$ spinons.[8] For $Cs_2CuBr_4$, AFM order exists below $T_N = 1.4$ K. By applying a magnetic field at around 30 T, there is a quantum phase transition into the FM state.[5]

$Cs_2CuCl_4$ and $Cs_2CuBr_4$ motivated the study of the magnetic properties of $Cs_2CuCl_{4-x}Br_x$. This mixed system allows studying the effects of controlled quenched disorder on the physics of the triangular lattice, because the preferential occupation of certain sites by either Cl or Br in the $[CuX_4]$ tetrahedra leads to a selective occupation.[9,10] The idea for the subsequent replacement of the three Cl crystallographic sites by Br has been obtained from the x dependence of the lattice parameters, which has shown anisotropy through an investigation using x-ray powder diffraction. The crystal structure *P*nma of all compositions remains unchanged over the whole concentration range.[10] At low temperatures, the crystal structures remain orthorhombic without any structural phase transition down to 20 K. Particularly, the anisotropy of the thermal expansion varies for different x, leading to distinct changes of the geometry of the local $[Cu^{2+}]$ environment as a function of the respective composition.[9] **Figure 2** shows, for example, the crystal structure for a Br concentration of x=1.

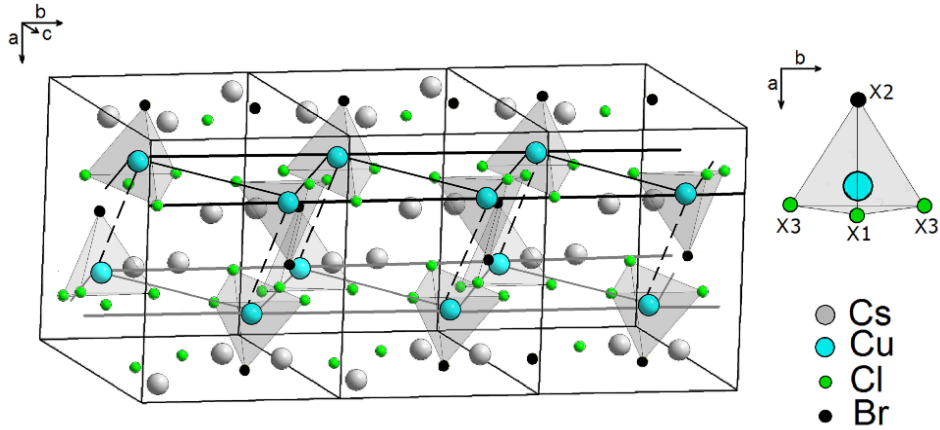

**Figure 2.** (left) crystal structure for x=1 with the $[CuCl_3Br_1]$ tetrahedra for three neighbouring unit cells in the bc-plane, and (right) schematic $[CuX_4]$ tetrahedra with Cl and Br having different preferred occupation of the X-crystallographic sites. For a Br concentration of x=1, the crystallographic position X2 is fully occupied.

For $Cs_2CuCl_4$, the magnetic structure (below $T_N$=0.62 K) is incommensurate along the chain direction with a temperature-independent ordering wave vector $q \approx 0.472$.[6] The magnetic structure is cycloidal with spins rotating in a plane that contains the b-axis in the unit cell. Coldea et al. described that the opposite sense of the spin-rotation in the chains (separated in a-direction) is due to an antiparallel orientation of their x- and y-spin-components.[6] For $Cs_2CuBr_4$, an incommensurate structure occurs below $T_N$=1.4 K with an ordering vector $q \approx 0.575$.[5]

The investigation of the ordering temperatures of the mixed system $Cs_2CuCl_{4-x}Br_x$ was performed with specific heat measurements in case of a larger x doping region. The ordering temperature $T_N$ decreases drastically with increasing Cl doping.[5]

A density functional theory (DFT) study of the microscopic properties of $Cs_2CuCl_4$ and $Cs_2CuBr_4$ was performed and compared with the experimental results.[11] A calculation with the exchange-correlation functional was made in order to investigate the dependence of electronic and magnetic properties on the used functional. The analysis of $J$ and $J'$ using the available x-ray structural data for $Cs_2CuCl_4$ and $Cs_2CuBr_4$ shows that the changes of the exchange couplings and the frustration coefficients depend on the geometry of the $[CuX_4]$ tetrahedra. The calculated ratio of $J'/J$ for $Cs_2CuCl_4$ is 0.384, being very close to the experimental result of 0.34.[7,11] For $Cs_2CuBr_4$, this ratio equals to 0.64, which is in good agreement with the experimental values of 0.74 and 0.41.[5,11,12]

In this paper, we present the results of a single crystal neutron diffraction study and DFT calculation of selected compositions of the $Cs_2CuCl_{4-x}Br_x$ mixed system. Section 2 shows the results of the experiments, the DFT calculations, and the discussions of the physical properties for the low-dimensional spin system $Cs_2CuCl_{4-x}Br_x$, followed by the conclusion and outlook. In Section 4 we provide experimental details.

## 2. Results and Discussions

Single crystal neutron diffraction has been used to prove the interrelation between the selective occupation of the X-sites in the tetrahedra and the magnetic properties. Scans through the (*0 q 0*) magnetic reflection with increasing temperature showed that the intensity of the magnetic reflection decreases, when the temperature approaches the Neel point $T_N$ (**Figure 3a**). For example, for the compound $Cs_2CuCl_3Br_1$ a well-defined magnetic peak was observed for *q* close to the half-integer value (0, 0.522(1), 0).

**Figure 3c** presents the *q*-scan for the three compositions $Cs_2CuCl_3Br_1$, $Cs_2CuCl_{2.6}Br_{1.4}$ and $Cs_2CuCl_{2.2}Br_{1.8}$. Whereby the intensity of the magnetic reflection of $Cs_2CuCl_{2.6}Br_{1.4}$ decreases in comparison to $Cs_2CuCl_3Br_1$, $Cs_2CuCl_{2.2}Br_{1.8}$ shows no magnetic reflection. **Figure 3d** presents the *k*-value for $q_y \approx 0.519$ of $Cs_2CuCl_{2.6}Br_{1.4}$, which changes slightly, depending on the temperature. The results, summarised in **Table I**, show the position of the (*0 q 0*) reflections for the investigated compositions.

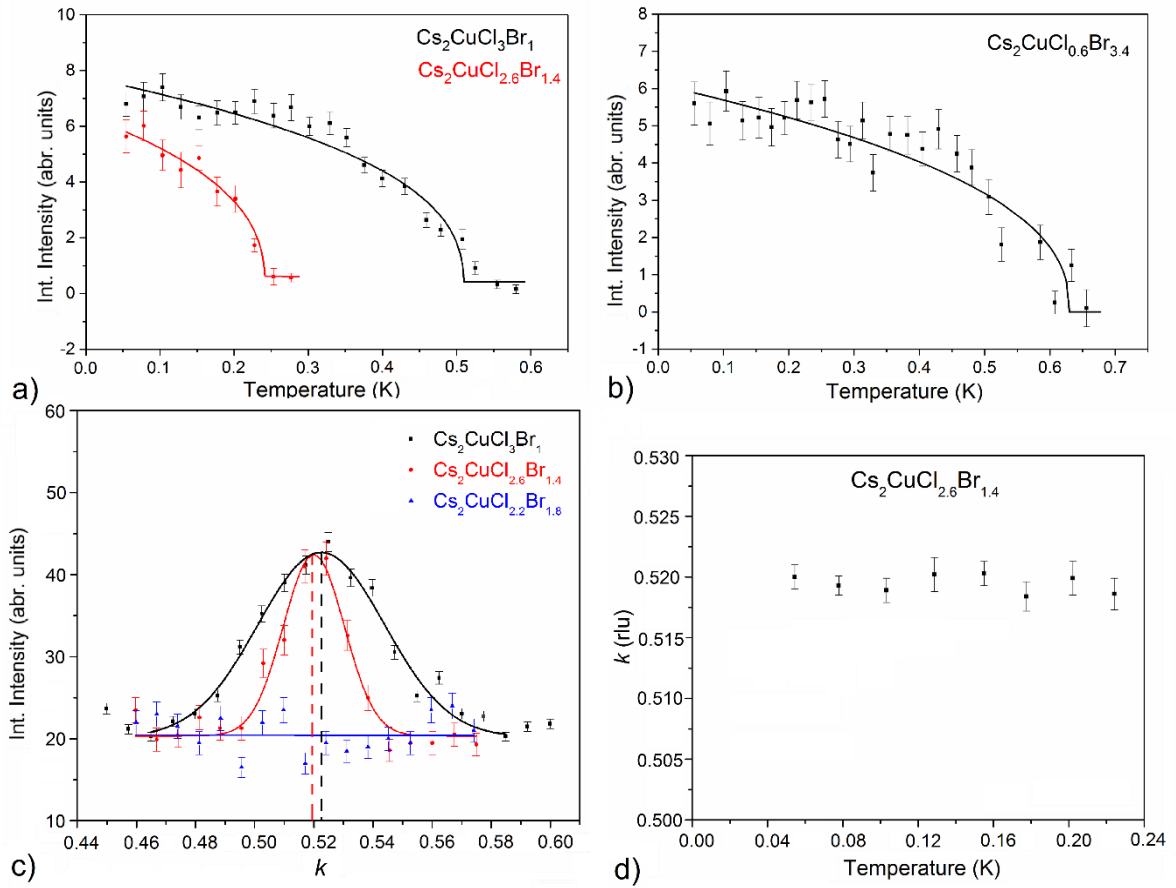

**Figure 3.** Temperature dependence of the intensity for ($0\ q\ 0$) peak of **a)** $Cs_2CuCl_3Br_1$ and $Cs_2CuCl_{2.6}Br_{1.4}$; for the scans the collimators 80` and 20` were used, respectively; and **b)** $Cs_2CuCl_{0.6}Br_{3.4}$; the collimator 40` was used, **c)** $q$-scan for $Cs_2CuCl_3Br_1$, $Cs_2CuCl_{2.6}Br_{1.4}$ and $Cs_2CuCl_{2.2}Br_{1.8}$ at 50mK; the intensity for the magnetic reflection of $Cs_2CuCl_3Br_1$ has been scaled down by a factor of 16, due to using different collimators; the two vertical dashed lines illustrate the shift of the ($0\ q\ 0$) peak positions, **d)** temperature dependence of the $k$-value of $Cs_2CuCl_{2.6}Br_{1.4}$.

**Table I.** Position of the ($0\ q\ 0$) reflections, the Neel temperature and the chemical compositions measured by EDX for $Cs_2CuCl_{4-x}Br_x$. The data for $Cs_2CuCl_4$ and $Cs_2CuBr_4$ are taken from literature.[5,6]

| Nominal chemical composition | EDX results | $q_y$ | $T_N$, [K] |
|---|---|---|---|
| $Cs_2CuCl_4$ [6] | - | (0 0.528(2) 0) | 0.62 |
| $Cs_2CuCl_3Br_1$ | x=0.82(7) | (0 0.522(1) 0) | 0.51(1) |
| $Cs_2CuCl_{2.9}Br_{1.1}$ | x=1.09(5) | (0 0.516(3) 0) | 0.34(1) |
| $Cs_2CuCl_{2.6}Br_{1.4}$ | x=1.38(6) | (0 0.519(1) 0) | 0.24(2) |
| $Cs_2CuCl_{2.2}Br_{1.8}$ | x=1.85(5) | not observed down to 50mK | |
| $Cs_2CuCl_{1.8}Br_{2.2}$ | x=2.15(6) | not observed down to 50mK | |
| $Cs_2CuCl_1Br_3$ | x=3.16(6) | not observed down to 50mK | |
| $Cs_2CuCl_{0.7}Br_{3.3}$ | x=3.33(5) | (0 0.560(4) 0) | 0.33(1) |
| $Cs_2CuCl_{0.6}Br_{3.4}$ | x=3.45(7) | (0 0.573(1) 0) | 0.63(1) |
| $Cs_2CuBr_4$ [5] | - | (0 0.575 0) | 1.4 |

Seeing the changes of the positions of $q_y$ for the investigated compounds, we expect that the compounds $Cs_2CuCl_3Br_1$ and $Cs_2CuCl_{2.6}Br_{1.4}$ have a similar cycloidal incommensurate magnetic structure as $Cs_2CuCl_4$. The position of $q_y$ of the $Cs_2CuCl_{0.6}Br_{3.4}$ compound is similar to $Cs_2CuBr_4$. This implies that they have the same incommensurate magnetic structure. During the single crystal neutron diffraction experiments, we measured up to six magnetic reflections for each composition. For the determination of the magnetic structure, the number of the reflections is not sufficient. Nevertheless, the observed reflections comply with previously published results.[5,6]

The results of the single crystal neutron diffraction experiment for this mixed system showed that the zero-field ordering temperature for small x varies between $T_N$ = 0.62 K (x = 0) and $T_N$ = 0.24 K (x = 1.4), see **Figure 4**. For the middle range between 1.5 < x < 3.2, by investigating two compositions with Br concentrations of x = 1.8 and x = 3, no magnetic peaks were found. In addition, we investigated the composition with a Br concentration of x = 2.2. The results show the absence of magnetic order in the investigated direction of the incommensurate wave vector $q$, which means that the magnetic order is completely suppressed down to at least 50 mK. The magnetic order may be suppressed by the two fundamentally different mechanisms: frustration (change of the frustration coefficient $\alpha = J'/J$) and dimensionality (preferred occupation of the halogen site mediating the interlayer exchange). In addition, disorder in such systems can also be a reason for a suppressed magnetic order.[13]

On the opposite side of the phase diagram, the ordering temperature is reduced from $T_N$ = 1.4 K (x = 4) to $T_N$ = 0.33 K in case of a slightly smaller x (x = 3.3). For $Cs_2CuCl_4$, the inelastic neutron scattering (INS) was used to explore the magnetic excitations. In the spin liquid phase above the 3D ordering transition, a broad continuum of excited states was observed.[8] Coldea et al. constitute that this is a two-spinon continua in the spin liquid phase above $T_N$, and that the spin liquid phase is a 2D spin liquid (2D SL). [8,14] The question, if this 2D SL exists for the part of the Br concentration range of this mixed system, is still to be confirmed with INS and, therefore, we have marked 2D SL in **Figure 4** with "?".

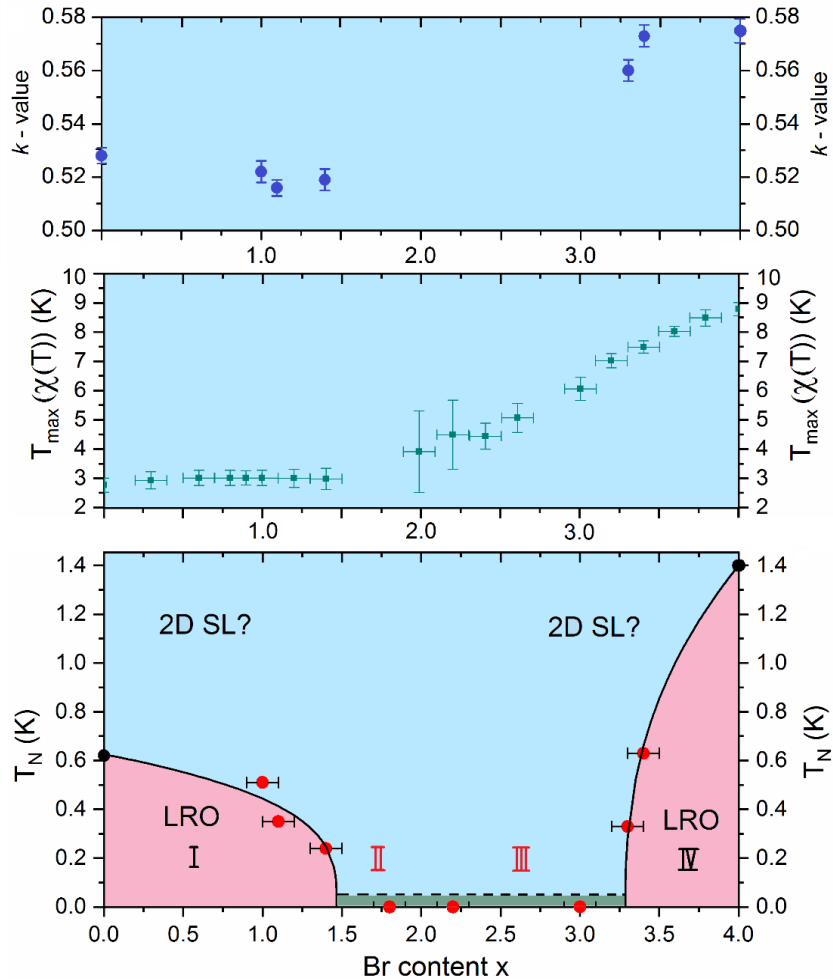

**Figure 4.** Magnetic phase diagram of $Cs_2CuCl_{4-x}Br_x$; the positions of the strongest magnetic reflections (*0 q 0*) at the top (blue circles) in regimes I and IV; the maxima in the magnetic susceptibility (green squares in the middle panel), published from Cong et al.[15]; at lower temperatures a long-range magnetic order (LRO) is observed for some compositions (red circles) in regimes I and IV, the black circles show the data from literature of $Cs_2CuCl_4$ and $Cs_2CuBr_4$ [5,6]; magnetic properties below 0.05 K are not yet determined; the horizontal arrow bars ±0.1 represent the uncertainties in the EDX results of the chemical composition.

The full concentration range of $Cs_2CuCl_{4-x}Br_x$ will be divided into four regimes: regime I attributed to the $0 < x \leq 1.5$ Br content, regime II for $1.5 < x \leq 2$, regime III for $2 < x \leq 3.2$, and regime IV for $3.2 < x < 4$ (see **Figure 4**, bottom). The sectioning of the phase diagram into 4 regimes is based on the observed magnetic order (regimes I and IV) and the sections without magnetic order (regimes II and III), with special attention to regime III with its altering frustration lattice model. Regime I is characterized by the $Cs_2CuCl_4$ - type magnetic order, which we observed in our single crystal neutron diffraction for compounds with a Br concentration x up to 1.5. The magnetic regime I, which was determined by Cong et al. up to x = 1, does not agree with our results for the magnetic order in this mixed system.[15] Nevertheless, our observation that the regime I is stretched out up to x = 1.5, corresponds with the description of the magnetic susceptibility results of Cong et al.[15] They expect that up to x (slightly larger than 1.4), $T_{max}$ is almost independent from the Br concentration (see **Figure 4**, middle), and that the value $\chi_{mol}(T_{max})$ reveals a distinct reduction with x.[15] The magnetic interaction in regimes II – IV will be discussed later.

The *k*-value demonstrates that there is a big difference between the two regimes I and IV (see **Figure 4**, top), which show a magnetic order, and, therefore, a smooth interpretation between x = 0 and x = 4 does not exist.

To understand the relationship between the experimental results from neutron scattering for the magnetic phase diagram and the theoretical model for interaction, it is important to compare the results of the real crystal structure and the structure, which is used for the theoretical calculations. The substitution of Cl by Br in the Cu-tetrahedra is realised on the three crystallographic positions X1, X2, X3 (see Figure 2). Depending on the Br-concentration, different crystallographic positions are occupied by Br. For 0<x<1, the preferred occupied crystallographic position is X2, for 1<x<2 the preferred crystallographic positions are X1, X2, and for 2<x<3 it is X3.[16] Each crystallographic position is occupied with a specific deviation.[16] That means that for x=0.8 the variance of the preferred occupation of the crystallographic position X2 is 4%.[16] For x=1 this value is 8% and rises with an increasing Br concentration (for x=2 - 12% and for x=3 - 16%).[16] The nuclear order by way of substituting Cl by Br in the Cu-tetrahedra is decreasing with increasing x. The structure analysis in Ref. 16 gives an overview about the preferred occupation with disorder in real crystals of selected compounds of the $Cs_2CuCl_{4-x}Br_x$ mixed system. For the theoretical calculations in this work, the ordered structure type for x=1 and 2 without any deviation was used. This system is divided into four regimes depending on x with respect to the magnetic order and the exchange paths. To understand the interaction in this system, it is important to execute a DFT calculation. The exchange parameters are used for the description of the exchange paths and the choice of the Heisenberg model on an anisotropic triangular lattice (ATL) and/or the Heisenberg model on a square lattice (SL). The DFT calculation was performed only for ordered compositions, because a DFT calculation for partial occupation of halogen positions in Cu-tetrahedra would be too complex.

We have executed DFT calculations of the exchange parameters in order to understand the exchange couplings in this system. The exchange couplings constants, calculated by DFT total energy difference calculations based on the SCAN MGGA approximation for $Cs_2CuCl_4$ and $Cs_2CuBr_4$ and compounds with different Br concentration, are shown in **Table II**. [11,17-20] The $J$ for two ordered compositions of x = 1 and 2 in this mixed system change from AFM to FM behaviour. For the same compositions, the $J'$ stay AFM and these values increase up to a Br concentration of x = 2. The values of $J''$ remain AFM and also increase from x = 0 to x = 4.

**Table II.** DFT calculation of exchange coupling constants $J$, $J'$ and $J''$ with the SCAN method for the investigated compounds.

| Compound | SCAN [meV] | | | |
|---|---|---|---|---|
| | $J$ | $J'$ | $J''$ | $|J'/J|$ *) |
| $Cs_2CuCl_4$ | 1.27 | 0.30 | -0.09 | 0.24 |
| Exp. value for $Cs_2CuCl_4$ [12] | $J$=0.374[8](0.41)[12] | $J'$=0.125[8] | $J''$=-0.017[8] | $J'/J$=0.34 [8] (0.3) |
| $Cs_2CuCl_3Br_1$ | -0.33 | 0.52 | -0.29 | 0.63 |
| $Cs_2CuCl_2Br_2$ | -0.41 | 1.10 | -0.51 | 0.37 |
| $Cs_2CuBr_4$ | 0.28 | 0.67 | -0.39 | 0.42 |
| Exp. value for $Cs_2CuBr_4$ | $J$=0.97[5] (1.28)[12] | $J'$=0.72[5] (0.52)[12] | $J''$=-0.02[12] | $J'/J$=0.74[5](0.41)[12] |

*) For the determination of the ratio for $J > J'$ we used $J'/J$ and for $J < J'$ - $J/J'$.[21] For this calculation, $J > 0$ and $J' > 0$ means that both couplings are antiferromagnetic, which introduces

frustration into the Heisenberg model. $J'/J$ regarding $Cs_2CuBr_4$ is estimated by Zvyagin et al. and Ono et al. by comparing ratios of various theoretical models.[5,12]

The results show for $Cs_2CuCl_4$ that the ratio is 0.24 for the calculation of the experimental structure. The experimental ratios 0.3 and 0.34 are close to our DFT calculation with SCAN.[8,12] In case of $Cs_2CuBr_4$, we have calculated the ratio resulting in 0.42. The experimental ratios 0.41 and 0.74 show broad intervals for $Cs_2CuBr_4$, and the results are in good agreement in this interval.[5,12]

In comparison to our DFT calculation, Foyetsova et al. obtained for the ratio of the calculated coupling exchange constants with 0.24 for $Cs_2CuCl_4$ the same result, with 0.64 for $Cs_2CuBr_4$ a slightly larger one, and, in addition, different absolute values for $J$, $J'$ and $J''$.[11] For $Cs_2CuBr_4$, the values for $J$ and $J'$ of Foyetsova et al. are opposed ($J$ =1.66 meV, $J'$ = 1.04 meV).[11] The values for $J''$ are larger in our calculation and, in addition, are ferromagnetic, in comparison to Foyetsova et al. ($J''$ = 0.09 meV).[11] Our results show that the layers in a-directions are stronger coupled.

To describe the exchange paths in this system, we use the Heisenberg model on an ATL in case of $J > J'$, and, in case of $J < J'$, the Heisenberg model on a SL.[21,22] The schematic presentations of ATL and SL are shown on the left top for $Cs_2CuCl_4$ and the right top for $Cs_2CuBr_4$ of **Figure 5**, respectively. In general, the AFM interaction with and without frustration seems to be inherent in this mixed system and can be explained by the two models. In regime I, an anisotropic triangular lattice with frustration changes from x = 0 to x = 1 into the square lattice model without frustration. According to our calculation, the AFM coupling $J$ in b-direction has changed from an AFM to a FM one. That implies, there is no frustration. The AFM behaviour and the value of $J'$ in the bc-plane remain almost the same in this Br concentration range. The magnetic order lasts unchanged, and this observation is based on the structure details, especially by the preferred occupation of the tetrahedra crystallographic positions X1 for a Br concentration x up to 1 (see **Figure 5**, bottom). That means that the triangular lattice with a strong square lattice coupling ($J'$) stays up to a Br concentration x = 1. For a Br concentration 1 < x < 1.5, the preferred occupation of the tetrahedra crystallographic positions is X2, because the crystallographic position X1 is fully occupied by Br for this concentration region. This results in an unsymmetrical interaction case of the anisotropic square lattice, for example $J'_2 = J'_3 \neq J'_4 = J'_5$, and can be considered as a coupling disorder. The magnetic order can be expected for x up to 1.4, with $T_N$ rapidly decreasing from 0.51 K for x = 1 to 0.24 K for x = 1.4.

In regime IV, the chain coupling constant $J$ decreases and changes from a FM to an AFM one and induces frustration in this concentration range (see **Figure 5**). $J'$ also decreases but maintains the same AFM behaviour. This regime is also marked by the magnetic order as a result of our neutron diffraction. Here, a preferred occupation of Br of the two equivalent X3 positions happens.[9,10]

In the following, we will discuss the model for the interaction of regimes II and III in this mixed system (**Figure 5**). For the design of this model, the behaviour of the exchange parameter, received from the DFT calculation, was used. In these regimes II and III, no magnetic order has been detected down to 50 mK (**Figure 4**). The schematic 2D layer structure has two different interactions in the *bc*-plane (the chains run along the *b*-axis and between the chains) in the range of the Br concentration $1.5 < x \leq 3.2$. The zig-zag coupling $J'$ will be changed (enlarged) with increasing x by the exchange path via the X1 tetrahedra position within regime II. As a result of x rising within regime III, the chain coupling $J$ will also change from FM to AFM due to the exchange path via the X3 position (**Figure 5**).

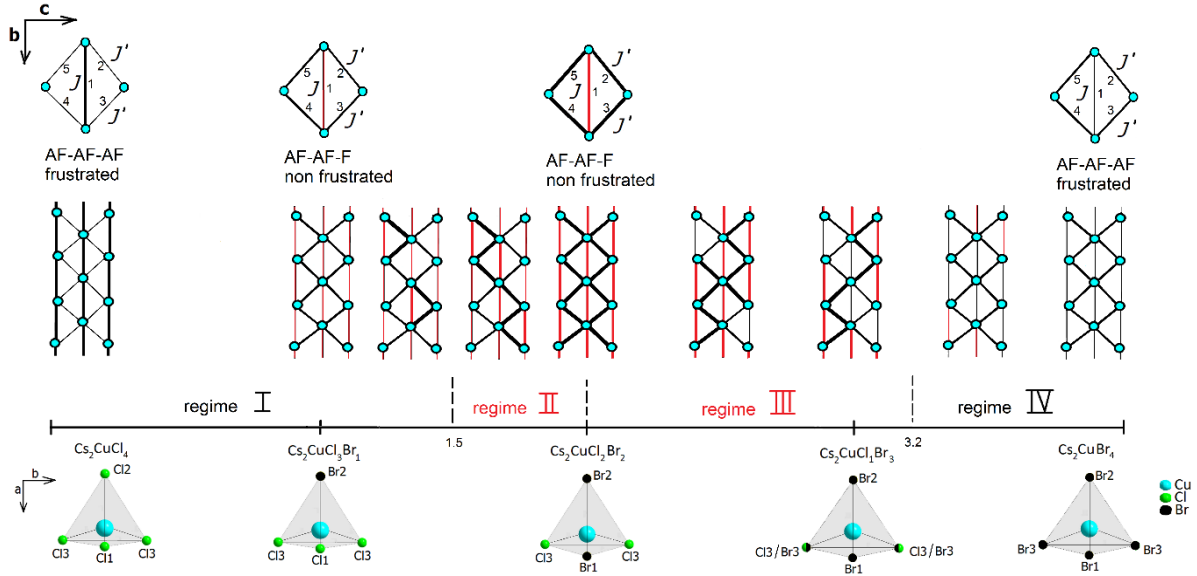

**Figure 5.** Controllable anisotropic triangular lattice in $Cs_2CuCl_{4-x}Br_x$: on the top, the model can also be viewed as a square lattice with an extra exchange along one diagonal for the ordered compositions $Cs_2CuCl_4$, $Cs_2CuCl_3Br_1$, $Cs_2CuCl_2Br_2$ and $Cs_2CuBr_4$; below, magnetic exchange paths in the *bc*-layer depending on the preferred Br-occupation.[9,10]

A special feature is the compound $Cs_2CuCl_2Br_2$, which shows a very strong square lattice coupling $J'$ and, in addition, a FM chain coupling $J$. In general, for this compound a magnetic order should be observable, but it could be that the structural disorder for a Br concentration of x = 2 leads to a coupling disorder, which results in an absence of any magnetic order. For the Br concentrations x = 1.8 and 2.2, there is no magnetic order in the investigated directions (0 1 0, 0 1 2, 0 1 -2).

In our opinion, the preferential occupation is the mechanism behind the controlled coupling models for a number of ordered Br concentrations.[9,10] The special behaviour of regime II is a modulation for the values of $J$ and $J'$ (exchange coupling along and between the chains) as a result of the occupation of the crystallographic position X2. The special behaviour of regime III is characterized by the modulation of the value of $J'$ (exchange coupling between the chains) and the change and control of the behaviour of $J$ (exchange coupling along the chains) through the other crystallographic position X3. For the ordered compositions with x = 1, 2, 4, which were also used for the DFT calculation, we present a triangular lattice with a square lattice coupling $J'$, perturbed by a chain coupling $J$ with a variously behaviour and different values (see **Figure 5**).

Summing up, this schematic 2D layer structure shows for this mixed system (including compounds with x=0 and 4) in the bc-plane two important interactions ($J$, $J'$). In general, for a non-frustrated lattice model, the magnetic order temperature should be higher than that for a frustrated lattice one. For example, in regime I up to x=1, the magnetic order temperature remains almost constant, because the structural disorder in real crystals counteracts or suppresses this order temperature. For compounds within regime I and x>1, the order temperature decreases rapidly, as the structural disorder becomes larger due to an increased Br concentration and as there are more available crystallographic positions in the Cu-tetrahedra for a Br occupation. In regimes II and III, the structural disorder increases. In these two regimes, no magnetic order can be measured up to 50mK. Regime III is characterized by the transition from a non-frustrated lattice model to a frustrated one. However, local lattices could be frustrated. Therefore, the exciting question of discovering a 2D spin-liquid is justified. And finally, regime IV (with x>3.2) is characterized by LRO.

Previous studies of $Cs_2CuBr_4$, using high-temperature series expansions by Zheng et al., resulted in $J''/J = 0.5$, but also report that a wide range of ratios $J''/J$ between 0.35 and 0.55 give comparable fits.[21] Another study extracted the Heisenberg couplings from high-field electron-spin-resonance measurements with the harmonic spin-wave theory.[12] Their findings correspond to $J''/J = 0.41(0)$. Nevertheless, the fore-mentioned coupling ratios are significantly higher than $J''/J = 0.3(2)$, obtained by Hehn et al., who have also discussed the case, which corresponds to a triangular lattice with a square lattice coupling and which has resulted in the ratio $J/J' = 0.2(1)$.[22] The ratio of exchange couplings for $Cs_2CuCl_4$ using high-temperature series expansions by Zheng et al, equal to $J''/J = 0.30(3)$, are in good agreement with the experimental results $J''/J = 0.34(3)$ from the neutron scattering experiments.[7,21] Using high-temperature series expansions, it is possible to refine/calculate the exchange coupling parameters of compounds with different Cl/Br concentrations (not only ordered compositions) and, subsequently, to compare $J$ and $J'$ with the experimental data of this mixed system.

## 3. Conclusion and outlook

In the triangular antiferromagnetic $Cs_2CuCl_{4-x}Br_x$ mixed system the changes in the $[Cu^{2+}]$ environment have a significant impact on the variations of the magnetic behaviour or the control of it. We present the novel magnetic phase diagram as a function of the Br concentration for this mixed system, which was determined by neutron diffraction. The magnetic phase diagram in zero magnetic field shows four different regimes.

Regimes I and IV correspond to different magnetic order, respectively. The first magnetic phase is described by the Br concentration x < 1.5 and the second one in regime IV for x > 3.2 (see **Figure 4**). In the magnetic phase diagram, the two ordered magnetic phases in regimes I and IV seem to be separated by the quantum critical points (QCP), $QCP_1$ near x = 1.5 and $QCP_2$ near x = 3.2, respectively. In literature, many mixed systems exist, which show such QCP between two magnetic phases depending on the doping concentration. [3,23,24]

In addition, measurements below 50 mK might unveil new results in the regimes II and III. Such measurements of physical behaviorat at very low temperatures have already been executed, for example down to 0.53 mK, as mentioned by Om Prakash et al.[25] We have presented a suggestion about the magnetic exchange paths in the bc-layer for the regimes II and III. The magnetic correlations can be understood through the structure features of the $[CuX_4]$ tetrahedra and the DFT calculation and by using the exchange-path model, which is presented in **Figure 5**.

This mixed system allows the study of changes from frustrated to non-frustrated behaviour by magnetic order in regime I and changes from non-frustrated to frustrated behaviour without magnetic order in regime III. Furthermore, specific heat measurements at very low temperatures (below 50 mK) of the same compositions from the regions II and III, which was measured with neutron diffraction, are helpful to clarify the existence of the magnetic ordering in these regions.

The regimes II and III are of high interest, to identify the spin Hamiltonian and the dominant spin correlation in these regimes. A quantitative analysis of the sharp lower boundary of the continuum and of the spin-wave-like dispersions can identify the coupling regimes. We are especially interested in the 2D spin-liquid in the region of $2 < x \leq 3.2$. Due to the strong sensitivity of the magnetic exchange on the structural details affecting the strength of interactions, we believe that $Cs_2CuCl_{4-x}Br_x$ is an exceptional model system, in which a number of "magic" compositions can be studied.

## 4. Experimental Section

The $Cs_2CuCl_{4-x}Br_x$ crystals with an orthorhombic structure (*P*nma) for the full Br concentration range were grown from aqueous solutions with the evaporation method. Single crystals were grown of nominal compositions with x = 0, 1, 1.1, 1.4, 1.8, 2.2, 3, 3.3, 3.4. The stoichiometric mixtures of the salts $CsCl/CsBr$ and $CuCl_2/CuBr_2$ were used for the growth.[10] For the crystal growth process, a small evaporation rate of 15mg/hour was applied, to reach the appropriate crystallization rate. Crystals generally grew within 6 - 8 months at a temperature of 50 °C.[16] Single crystals are shown in **Figure 6**.

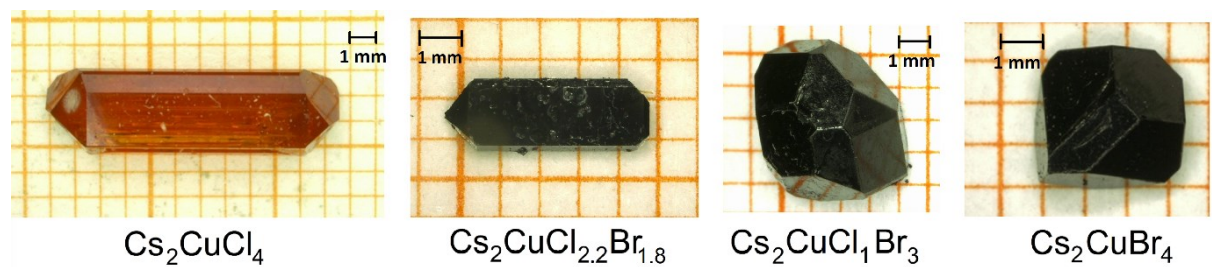

**Figure 6.** Single crystals of four compounds of the $Cs_2CuCl_{4-x}Br_x$ mixed system.

The chemical compositions of the samples were determined via energy dispersive X-ray analysis (EDX), using a electron microscope Zeiss DSM 940A. They are in good agreement with the nominal chemical composition (**Table I**). In this paper, we will use only the nominal chemical composition to name them.

Single crystal neutron diffraction experiments were carried out on the thermal-neutron diffractometer ZEBRA(TRICS) at SINQ of Paul Scherrer Institute (PSI) in Villigen, Switzerland.[26,27] For these experiments, single crystals (see above) with sizes around 4×6×2.5 mm³ were used. A monochromatic neutron beam with a wavelength of 2.317 Å was produced through a $PG_{002}$-monochromator and a PG-filter. Normal beam geometry was used. The diffracted reflections have been collected with a single ³He –tube detector. In front of the tube detector 80`, 40` or 20` soller collimators were installed to reduce background and thus to increase the peak-to-background ratio. Single crystal neutron diffraction was also performed on the cold three axes spectrometer MIRA at FRM II of Heinz Maier-Leibnitz Zentrum (MLZ) in Garching, Germany.[28] For the elastic measurements, a wavelength of 4.488 Å ($k_i$ = 1.4 Å⁻¹) was used by means of a PG monochromator. A Be-filter and a single ³He-tube detector were used. For both single crystal neutron diffraction measurements, the crystals were cooled down to 50 mK with a dilution.

The magnetic susceptibility of the single crystals of $Cs_2CuCl_{4-x}Br_x$ was measured at temperatures from 1.8 K to 300 K employing a Quantum Design PPMS/SQUID magnetometer. The samples were oriented with the b-axis along the magnetic field. The values of $T_{max}$ and $\chi_{mol}(T_{max})$ were found to be similar in all aspects to the results of the magnetic susceptibility reported by Cong et al., confirming the quality and compositions of the present single crystals.[15]

DFT calculations were performed with the DMol$^3$ code using a 4x4x4 mesh of k-points with its standard DNP basis set.[17,18] We use a meta GGA (MGGA) functional, which was introduced by Sun et al. as SCAN functional.[19,20] The exchange coupling integrals $J_{ij}$ of the Heisenberg Hamiltonian

$$H = \sum_{<i,j>} J_{ij} S_i S_j \quad (1)$$

can be obtained by means of DFT total energy calculations for an antiferromagnetic spin configuration on a 1x2x1 magnetic supercell, which contains eight Cu atoms, following the approach by Foyevtsova et al. For the computation, we used the experimental crystal structures at 20 K with ordered occupation of Cl/Br in the [CuX$_4$] tetrahedra.[9]

**Figure 7a** shows the interaction pathways $J$, $J'$, $J''$, $J_1$, $J_2$, $J_3$, $J_4$ and $J_5$ and the labelling of the Cu atoms in the 1x2x1 supercell of the $Cs_2CuCl_{4-x}Br_x$ mixed system. As the couplings $J''$ and $J_2$ are not separated in our calculation, they will be considered as a sum of both and named in the following only as $J''$.[11] The seven antiferromagnetic spin configurations $J_1$, $J_2$, $J_3$, $J_4$, $J_5$ by the total energy difference method, considered in order to calculate $J$, $J'$, $J''$, are presented in **Figure 7b**. In addition, a ferromagnetic spin configuration is also calculated.

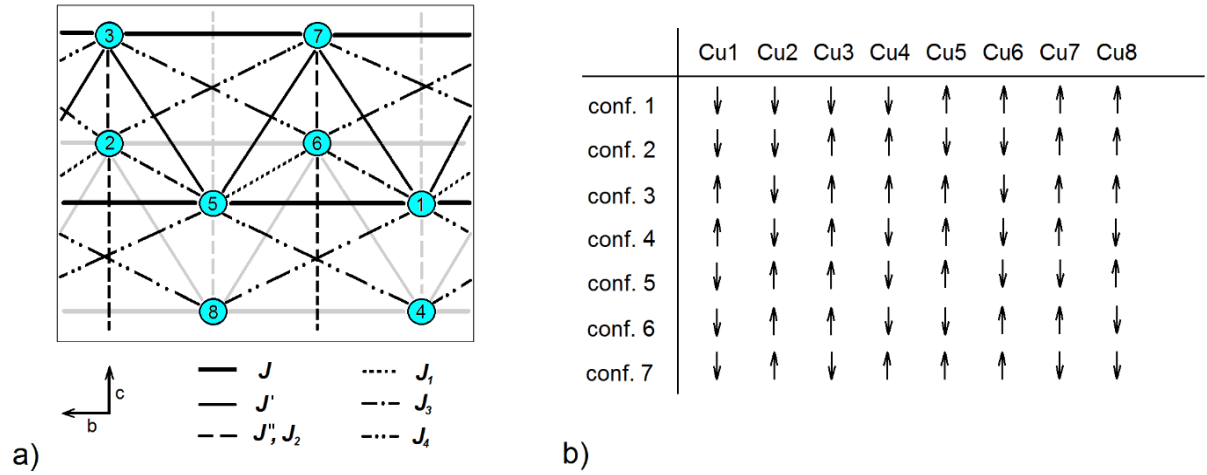

**Figure 7. a)** Schematic triangular lattice of Cu spins of the $Cs_2CuCl_{4-x}Br_x$ system and labelling of the Cu atoms in the 1x2x1 supercell with interaction pathways $J$, $J'$, $J''$, $J_1$, $J_3$, $J_4$, whereas $J_5$ is not shown, which interacts in a-direction, **b)** The seven selected spin configurations

Each of these spin configurations leads to a corresponding equation, from which we calculate the values of the exchange coupling constants. The energy difference ($\Delta E$) can be interpreted as $\Delta E = E_{FM} - E_{AFM}$, with $E_{FM}$ being the energy of the supercell in the ferromagnetic configuration of the Cu spins, and $E_{AFM}$ being the energy of the supercell in the antiferromagnetic configuration.

**Acknowledgements**

The authors thank W. Asmuss, F. Ritter, P. T. Cong, B. Wolf, M. Lang, C. Krellner and R. Valentí from Goethe-University, Frankfurt am Main and K. Foyevtsova from the University of British Columbia, Vancouver for fruitful discussions. The authors thank M. Medarde and T. Shang from Laboratory for Scientific Developments and Novel Materials (PSI), Villigen for their support during the susceptibility measurements. The neutron diffraction experiments were performed on ZEBRA(TRICS) at SINQ of PSI, Villigen, Switzerland and MIRA at FRM II of MLZ in Garching, Germany. This work was supported by Paul Scherrer Institut, Heinz Maier-Leibnitz Zentrum and Physic Department E21, Technische Universität München, University of Bayreuth, and the Deutsche Forschungsgemeinschaft through the research fellowship for the project WE-5803/1-1 and WE-5803/2-1.